# Tuning Excitonic Properties and Charge Carrier Dynamics by Halide Alloying in $Cs_3Bi_2(Br_{1-x}I_x)_9$ Semiconductors

*He Zhao\*, and Eline M. Hutter\**

H. Zhao, E. M. Hutter

Institute for Sustainable and Circular Chemistry, Department of Chemistry, Faculty of Science, Utrecht University, Universiteitsweg 99, 3584CG Utrecht, The Netherlands

\* H. Zhao: hezhao711603@outlook.com

\* E.M. Hutter: e.m.hutter@uu.nl

Funding: This work was supported by the Advanced Research Center Chemical Building Blocks Consortium (ARC CBBC) and Open Competition ENW-XS (grant OCENW.XS24.2.029).

**Abstract**

The perovskite-inspired bismuth halide semiconductor $Cs_3Bi_2Br_9$ is widely investigated as photoactive material for light-conversion applications. However, charge generation and separation are inherently limited by its modest sunlight absorption and strong exciton binding energy, respectively. Here, we demonstrate that both the light absorption and exciton dissociation are improved by controlled substitution of $Br^−$ with $I^−$ via mechanochemical synthesis of $Cs_3Bi_2(Br_{1-x}I_x)_9$. X-ray diffraction and Raman analyses confirm atomic-level halide mixing and reveal a crystallographic phase transition near $x = 0.8$. From absorption measurements on thin films, we determine the absorption coefficient, Urbach tail, and exciton binding energy for several $Cs_3Bi_2(Br_{1-x}I_x)_9$ compositions. From here, we find that the band gap can be tuned from 2.59 to 1.93 eV (for $x = 0.9$), while exciton binding energies reach a minimum at $x = 0.6$. Finally, transient absorption spectroscopy measurements suggest a weak correlation between recombination lifetime and Urbach energy, where the longest lifetimes are observed for the materials with lowest disorder. These results offer valuable insights for designing stable bismuth halide semiconductors with favorable light absorption properties and charge carrier dynamics.

# 1. Introduction

Perovskite-inspired bismuth halide semiconductors such as $A_3Bi_2X_9$, (A = $MA^+$ or $Cs^+$, X = $Br^-$ or $I^-$) have received increased attention as photoactive materials in photodetectors,[1] light-emitting diodes,[2] photocatalysis[3,4] and (indoor) photovoltaics[5,6] due to their low toxicity and high chemical and environmental stability under ambient conditions.[7] The widely used $Cs_3Bi_2Br_9$ has a trigonal crystal structure with a space group of $P\bar{3}m1$ and has shown promising performance as a light absorber in photodetectors and in photocatalysis.[7] However, its large band gap (ca. 2.6 eV) and high exciton binding energy are limiting factors for efficient light conversion, as these result in low light absorption and limited exciton dissociation into free carriers.[8,9] Halide alloying is a well-known method to tune the band gap and other photophysical properties of halide semiconductors. For instance, Yu et al. reported a series of $Cs_3Bi_2(Br_{1-x}I_x)_9$ ($0 \leq x \leq 1$) thin films with spin coating to modulate the band gap and crystallinity for photovoltaic applications, in which the $Cs_3Bi_2I_6Br_3$ thin films shows high crystallinity and compactness, reaching a power conversion efficiency of 1.15%.[10] Bonomi et al. utilized a radio frequency (RF)-magnetron sputtering method to prepare high-quality, single-phase $Cs_3Bi_2(Br_{1-x}I_x)_9$ films except for $x = 0.6$.[11] However, Hodgkins et al. suggested that single-phase powders were achievable up to a maximum of $x = 0.66$ (i.e., $Cs_3Bi_2I_6Br_3$) with solution method.[12] These inconsistent results motivate us to provide a detailed structure analysis of halide-alloyed $Cs_3Bi_2(Br_{1-x}I_x)_9$ semiconductors. In addition, the effect of halide alloying on excitonic properties and charge carrier dynamics within these materials still lags behind.

Herein, we use mechanochemical ball mill synthesis to prepare halide-alloyed $Cs_3Bi_2(Br_{1-x}I_x)_9$ semiconductors. This solvent-free synthesis technique gives us high control over the halide mixing ratio in the final compound,[13–15] because it circumvents solubility issues of precursors or reaction intermediates. The mixed-halide powders obtained with mechanochemical synthesis enable reliable phase identification through powder-based crystal structure analysis, which offers stronger diffraction signals and reduced orientation effects compared to thin films. Hence, we demonstrate that the maximum substitution of I into $Cs_3Bi_2Br_9$ is $x \sim 0.8$ by keeping the same $P\bar{3}m1$ phase of $Cs_3Bi_2Br_9$. Upon further increasing the I content to $x = 0.9$, a homogeneous phase forms with the same crystal structure of $Cs_3Bi_2I_9$ ($P6_3/mmc$). Raman spectra further confirm the formation of solid solutions for mixed-halide samples except for the I-rich compound at $x = 0.8$, where both phases are observed. We then prepared thin films for pure halide and mixed-halide samples through a solution-processed spin coating method and observe prominent excitonic peaks in absorption spectra. We find that the halide mixing ratio tunes both the band gap value and exciton binding energy of the mixed-halide $Cs_3Bi_2(Br_{1-x}I_x)_9$ materials, where the iodide-rich compositions have lower band gaps and lower exciton binding energies. Finally, transient absorption (TA) measurements were performed to reveal the effect of halide alloying on the charge carrier dynamics and TA analysis shows the longest charge carrier lifetimes for $x = 0.6$. Our findings provide new insights into the tunability of band gap, exciton binding energy and carrier decay lifetimes with halide alloying in bismuth halide semiconductors.

## 2. Results and Discussion

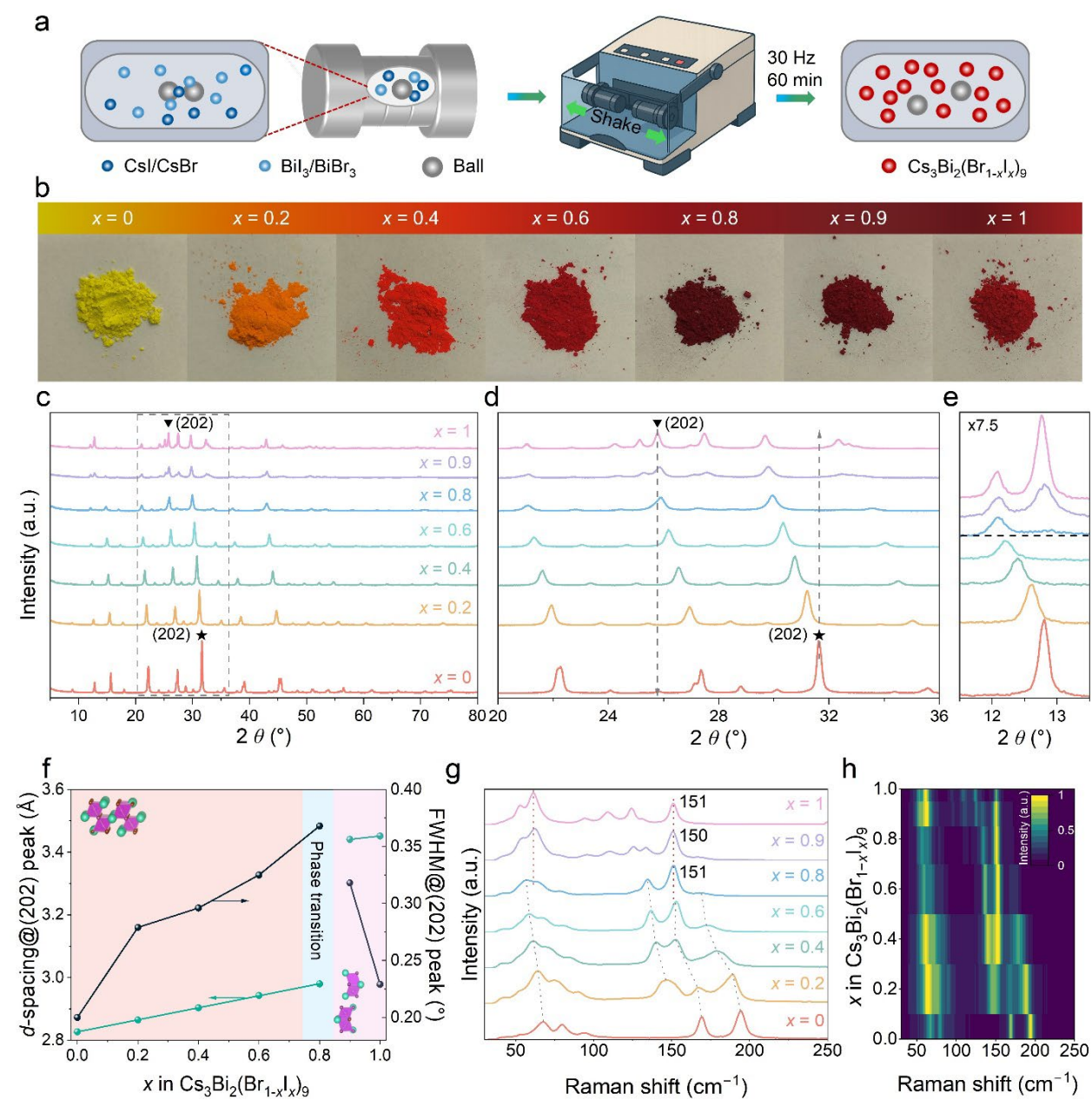

**Figure 1**. Mechanochemical synthesis and structural properties of $Cs_3Bi_2(Br_{1-x}I_x)_9$ powders. a) Schematic illustration of the preparation of $Cs_3Bi_2(Br_{1-x}I_x)_9$ powders by mechanochemical

synthesis. b) Optical photographs of the $Cs_3Bi_2(Br_{1-x}I_x)_9$ powders with different I content. c) XRD patterns of $Cs_3Bi_2(Br_{1-x}I_x)_9$ powders obtained from mechanochemical synthesis. Enlarged view in a $2\theta$ range of d) 20−36° and e) 11.5−13.5°. f) d-spacing and full width at half maximum (FWHM) derived from (202) peaks in XRD patterns as a function of the relative I content $x$. g) Raman spectra and h) the corresponding contour map.

The halide-alloyed $Cs_3Bi_2(Br_{1-x}I_x)_9$ powders were synthesized with mechanochemical synthesis approach as previously reported.[14,15] Briefly, the starting materials (CsI, CsBr, $BiI_3$ and $BiBr_3$, ca. 1 g in total) with different stoichiometric ratios are filled into the ball mill jars, together with three zirconia balls (ca. 10 g in total), and sealed in an inert glovebox, followed by horizontal shaking at room temperature (**Figure 1**a). As the milling jar oscillates horizontally, the movement activates the mixtures by the dispersion of solids, leading to the plastic deformation of the starting materials. This process causes generation, migration of defects in solids and creates contacts for the starting materials, thus resulting in chemical reactions and the formation of final products. By changing the ratio of starting materials, we were able to synthesize a series of $Cs_3Bi_2(Br_{1-x}I_x)_9$ ($x$ = 0–1) powders. From Figure 1b, we observed a continuous color change from yellow to dark red with the increasing I content up to 0.9, while the color turns to red subsequently at $x$ = 1. We first characterized the phase structure of the obtained samples with X-ray diffraction (XRD) crystallography (X-ray source: 1.5406 Å). Figure 1c displays the XRD patterns of $Cs_3Bi_2Br_9$, $Cs_3Bi_2I_9$, and mixed-halide samples. $Cs_3Bi_2Br_9$ crystalizes in a trigonal space group $P\bar{3}m1$ while the $Cs_3Bi_2I_9$ exhibits a hexagonal phase with space group $P6_3/mmc$, matching well with the simulation data (Figure S1) and excluding the presence of crystalline side phases. We also observe variation in the diffraction peak between full bromide and full iodide in the magnified (202) peak (Figure 1d). By substituting the smaller-size $Br^-$ (1.96 Å) with larger $I^-$ (2.20 Å),[16] the d-spacing of the (202) peak increases progressively, as suggested by the shift of diffraction peaks to lower diffraction angles. The gradual shift of XRD peaks instead of the coexistence of separate peaks belonging to $Cs_3Bi_2Br_9$ and $Cs_3Bi_2I_9$, verifies the formation of solid solutions with intermediate lattice parameters without phase segregation.[17] Figure 1f shows a linear relationship of d-spacing value with I ratio in the range between $x$ = 0 and $x$ = 0.8, indicating the nature of homogeneous alloying[18] while the powders exhibit reduced crystallinity from $x$ = 0 to $x$ = 0.8 as determined by the increased full width at half maximum (FWHM) of the (202) peak.[19] Another observation is that samples with $x \leq 0.8$ crystallize in a trigonal phase, similar to the full bromide, while for higher $x$ (> 0.8), the XRD pattern shows the typical $P6_3/mmc$ phase of Cs$_3$Bi$_2$I$_9$. At $x$ = 0.8, it shows a trigonal phase but has an additional small peak at 12.8° from I-rich phase (Cs$_3$Bi$_2$I$_9$), suggesting the formation of mixed solid solutions (Figure 1e).

To further verify the homogeneous alloying in the perovskite-inspired materials, Raman spectra for Cs$_3$Bi$_2$(Br$_{1-x}$I$_x$)$_9$ powders at room temperature were used to discern the local changes in octahedra. Figure 1g shows the evolution of full bromide, full iodide and halide-alloyed powders using a 785 nm laser in the range of 30–250 cm$^{-1}$. Cs$_3$Bi$_2$Br$_9$ and Cs$_3$Bi$_2$I$_9$ show a series of 7 peaks, with the strongest peaks near 67 cm$^{-1}$, 80 cm$^{-1}$, 94 cm$^{-1}$, 169 cm$^{-1}$ (E$_g$), and 194 cm$^{-1}$ (A$_{1g}$) in Cs$_3$Bi$_2$Br$_9$. The two strong peaks of A$_{1g}$ and E$_g$ are assigned to the in-plane and out-of-plane vibrational modes of the [BiBr$_6$]$^{3-}$ octahedron.[20] These peaks appear on the lower energy side for Cs$_3$Bi$_2$I$_9$ with Raman modes at 61 cm$^{-1}$ (E"; Bi–I bending mode), 94 cm$^{-1}$ (E"; asymmetric stretching), 109 cm$^{-1}$ (A'$_1$, symmetric stretching), 124 cm$^{-1}$ (E$_{2g}$; asymmetric stretching), and 150 cm$^{-1}$ (A$_{1g}$; symmetric stretching). The two modes at 94 cm$^{-1}$ (E$_{2g}$, asymmetric) and 109 cm$^{-1}$ (A$_{1g}$; symmetric) can be assigned to the bridged Bi–I stretching while another two peaks at 124 cm$^{-1}$ (E$_{1g}$; asymmetric) and 150 cm$^{-1}$ (A$_{1g}$; symmetric) are due to the terminal Bi–I stretching in the [Bi$_2$I$_9$]$^{3-}$ anion. These frequencies are in good agreement with those previously reported.[21,22] From the Raman spectra, it is evident that I mixing at the Br-site influences the octahedral vibrations in terms of vibrational frequencies and intensities. In halide-alloyed compounds, the A$_{1g}$ mode of Cs$_3$Bi$_2$Br$_9$ shows a gradual move towards a lower wavenumber and becomes less sharp (dotted lines in Figure 1g) due to the introduction of heavier I atoms and the longer and weaker Bi–I bonding strength than that of Bi–Br. A new peak located at around 147 cm$^{-1}$ is observed in the spectrum for the sample with $x$ = 0.2 and shows a clear red shift with increasing I content to 0.6. The continuous variations in the Raman peaks generally suggest the alloying nature (Figure 1h).[23] On the contrary, the alloy with $x$ = 0.8 shows the A$_{1g}$ modes of Cs$_3$Bi$_2$I$_9$ without considerable shift. This hints towards the formation of I-rich octahedra. However, at $x$ = 0.9, we observe a minimal shift of the Raman peaks compared to Cs$_3$Bi$_2$I$_9$, suggesting the formation of alloyed compounds. These results agree with the XRD analysis, suggesting that a homogeneous solid is obtained for $x$ = 0.2–0.6 and $x$ = 0.9 and a mixture of phases is formed for $x$ = 0.8.

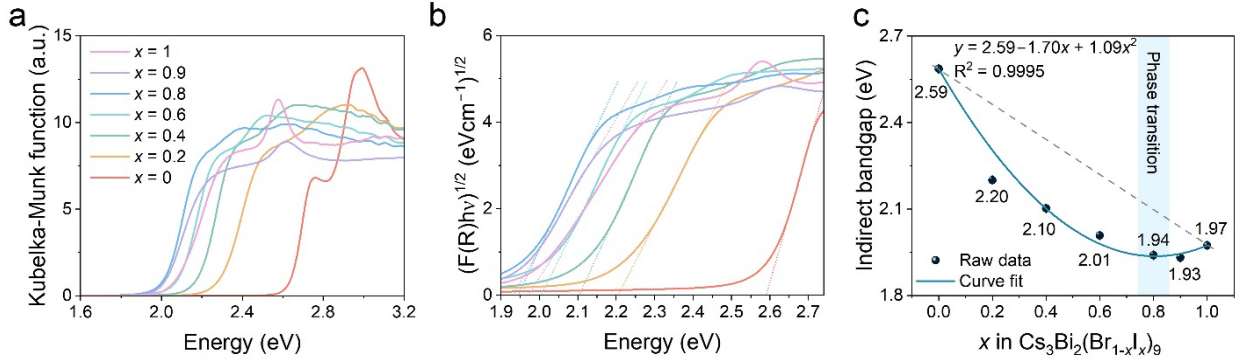

**Figure 2.** Optical properties of $Cs_3Bi_2(Br_{1−x}I_x)_9$ powders. a) Kubelka–Munk transformation of diffuse reflectance spectra. b) Tauc plots based on the Kubelka–Munk absorbance of $Cs_3Bi_2(Br_{1−x}I_x)_9$ powders. c) Indirect band gap versus the relative I content ($x$), estimated from the Tauc plot, together with the fitted bowing curve.

To investigate the optical properties of $Cs_3Bi_2(Br_{1−x}I_x)_9$ compounds, we measured the diffuse reflectance spectra of these powder materials at room temperature using an integrating sphere. **Figure 2**a-b show the Kubelka-Munk absorbance and Tauc plots of the absorption onsets, respectively. The $Cs_3Bi_2Br_9$ exhibits the largest optical indirect band gap of 2.59 eV, consistent with previously reported values,[24] while for the $Cs_3Bi_2I_9$ powder the indirect band gap occurs at 1.97 eV, close to values reported for macroscopic crystals (1.99 eV).[25] The absorption onset shows a progressive red shift when going from $x = 0$ to $x = 0.8$. The indirect band gap calculated for all of the powders and the simulated bowing curves are shown in Figure 2c. With 20% of I substitution, the band gap red-shifts toward 2.20 eV. A further increase in the I concentration reduces the band gap down to 1.94 eV (for $x = 0.8$). Notably, in the $Cs_3Bi_2(Br_{1−x}I_x)_9$ series, the lowest band gaps are observed for iodide-rich samples with $x = 0.8$, and 0.9, with optical band gaps of 1.94, and 1.93 eV, respectively, slightly lower than for $Cs_3Bi_2I_9$. We also notice that the band gap does not follow a linear trend between the band gap of full bromide and full iodide, but instead shows band gap bowing. The $Cs_3Bi_2(Br_{1−x}I_x)_9$ powders display band gap bowing with the bowing parameter $b$ of 1.09 for indirect band gaps, following the quadratic fitting rule $E_G(x) = xE_{G(x=1)} + (1-x)E_{G(x=0)} - bx(1-x)$ with $E_G$ the band gap value from the Tauc plot.

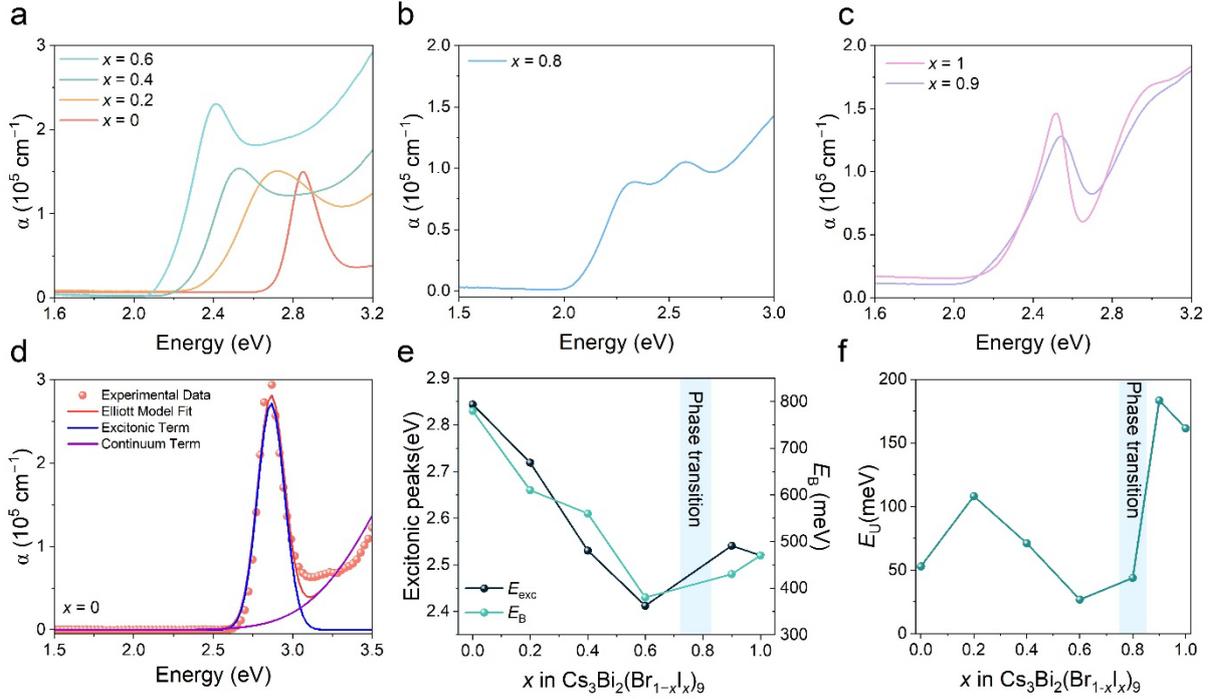

**Figure 3**: Absorption coefficients (α), exciton binding energy and Urbach energy ($E_u$) of $Cs_3Bi_2(Br_{1-x}I_x)_9$ thin films. a) Absorption coefficient for samples with a trigonal structure ($x = 0-0.6$). b) Absorption coefficient for $x = 0.8$, showing two excitonic peaks, indicating multiple phases with different halide mixing ratios. c) Absorption coefficient for samples with a hexagonal phase structure ($x = 0.9-1$). d) Absorption coefficient (salmon balls) of a $Cs_3Bi_2Br_9$ thin film fitted by Elliott's model (red line) with the contribution from excitonic (blue line) and continuum band (purple line) transitions. e) Plots of the change in excitonic peaks and exciton binding energy calculated from Elliott model fit as a function of the relative I content ($x$). f) Urbach energy extracted from the absorbance of $Cs_3Bi_2(Br_{1-x}I_x)_9$ thin films as a function of the relative I content ($x$).

To study the exciton and charge carrier dynamics of the synthesized materials, we used the products of the mechanochemical synthesis to fabricate thin films of $Cs_3Bi_2(Br_{1-x}I_x)_9$ through spin coating with antisolvent treatment, followed by ageing and annealing process (see details in section 4). The typical optical images of $Cs_3Bi_2(Br_{1-x}I_x)_9$ thin films are shown in Figure S2. Figure S3a displays the XRD patterns of the spin-coated thin films, showing the characteristic reflections of the alloyed $Cs_3Bi_2(Br_{1-x}I_x)_9$ without impurities. In addition, we observe that the diffraction peaks of (00l) family are relatively strong compared with the other indexed reflections (e.g., (202) observed in powder reference), suggesting the thin films most likely showed preferred growth along [001].[26] Note that the diffraction peak at 12.8° (from $Cs_3Bi_2I_9$) for $x = 0.8$ (Figure S3b) is more pronounced in the films than in the powders, which mostly likely originate from the ion migration and segregation at higher annealing temperatures (*i.e.*, 250 °) during spin coating.[27] Raman analysis of the thin films demonstrates the gradual shifts of Raman peaks observed in the corresponding powders (Figure S3c). Altogether, we demonstrate that the $Cs_3Bi_2(Br_{1-x}I_x)_9$ thin films can be successfully prepared with the powders from mechanochemical synthesis. We then measured the reflection and transmission on the thin films by UV-vis spectroscopy with an integrating sphere. The thickness of $Cs_3Bi_2(Br_{1-x}I_x)_9$ thin films were estimated by cross-sectional scanning electron microscopy and varies between 100-220 nm (Figure S4). On basis of the reflection, transmission and thickness of the thin films, we calculated the absorption coefficients, see **Figure 3**a-c and Figure S5. Both the excitonic absorption peaks and the absorption coefficients (i.e. $10^5$ cm$^{-1}$) of the mixed-halide thin films are close to values reported for full Br and full I single crystals.[9,18,28] Halide-alloyed films with $x = 0.2–0.6$ and $x = 0.9$ show single exciton peaks, suggesting the homogeneous mixing of bromide and iodide ions in this regime.[29] In contrast, halide-alloyed thin films for $x = 0.8$ shows two excitonic absorption bands, indicating the coexistence of two distinct phases with different mixing ratios, consistent with XRD and Raman results.

To understand the effect of halide alloying on the exciton binding energy ($E_B$) in $Cs_3Bi_2(Br_{1-x}I_x)_9$ compounds, we need to accurately separate the contributions to the absorption from continuum states and from Coulombically bound electron-hole pairs (excitonic states) because the excitonic states significantly influence the absorption coefficients near the band gap $E_G$. Here, we used the Elliott model consisting of continuum state absorption and multiple excitonic states (n ≤ 3).[9] The cumulative fit is in good agreement with experimental data as shown in Figure 3d and Figure S6. As shown in Figure 3e and Table S1, we determine the value of $E_B$ for $Cs_3Bi_2Br_9$ and $Cs_3Bi_2I_9$ to be 780 meV and 470 meV, respectively. Note that the extracted values are different from those reported data in powders, or single crystals[9,30] due to the strong structure-dependency of exciton binding energy.[31] In addition, our results provide new insights by demonstrating a continuous tunability of $E_B$ within the $P\bar{3}m1$ phase across the halide-alloyed compositions.[15,32,33] In particular, the lowest $E_B$ is observed at $x = 0.6$, reduced by 400 meV, indicating a strong compositional dependence of exciton binding energy for bismuth halide semiconductors, which has not been addressed before. This systematic decrease in $E_B$ indicates more efficient charge separation with the increase of I content up to the optimal ratio ($x = 0.6$). Note that we could not extract the exciton binding energy of $Cs_3Bi_2(Br_{0.2}I_{0.8})_9$ thin films due to the formation of a mixed solid solution.

We further derive the Urbach energies of the $Cs_3Bi_2(Br_{1-x}I_x)_9$ films from $\alpha = \alpha_0 \exp(h\nu/E_U)$, where α is the absorption coefficient, $\alpha_0$ is a constant and $E_U$ denotes Urbach energy.[34] The full bromide composition ($x = 0$) has an $E_U$ of around 53 meV which almost doubles (108 meV) for an iodide ratio of $x$ = 0.2 (Figure 3f), which is likely due to the lattice distortion resulting from iodide incorporation.[35] We note that the $x$ = 0.6 thin film shows the lowest $E_U$ among this series, down to around 27 meV, which is probably owing to the higher film homogeneity and stability at $x$ = 0.6.[10,36] The $E_U$ increases at $x$ = 0.8, where the crystal structure transforms from trigonal to hexagonal and increases more rapidly with $x$ > 0.8. This may originate from the widely observed irregular surface and poor surface coverage for I-rich cesium bismuth halides[5,37] and thus the higher Urbach energy.[38] A lower $E_U$ for $x$ = 0.6 composition indicates the less electronic disorder in the crystal lattice and better optoelectronic qualities.[39] These results coupled with the concurrent reduction of $E_G$ and $E_B$ at $x$ = 0.6 suggest that this composition is especially promising for light conversion applications, as it has a broader absorption of the solar spectrum, facilitates easier exciton separation into free charges and shows less energetic disorder, thereby improving the energy conversion efficiency.

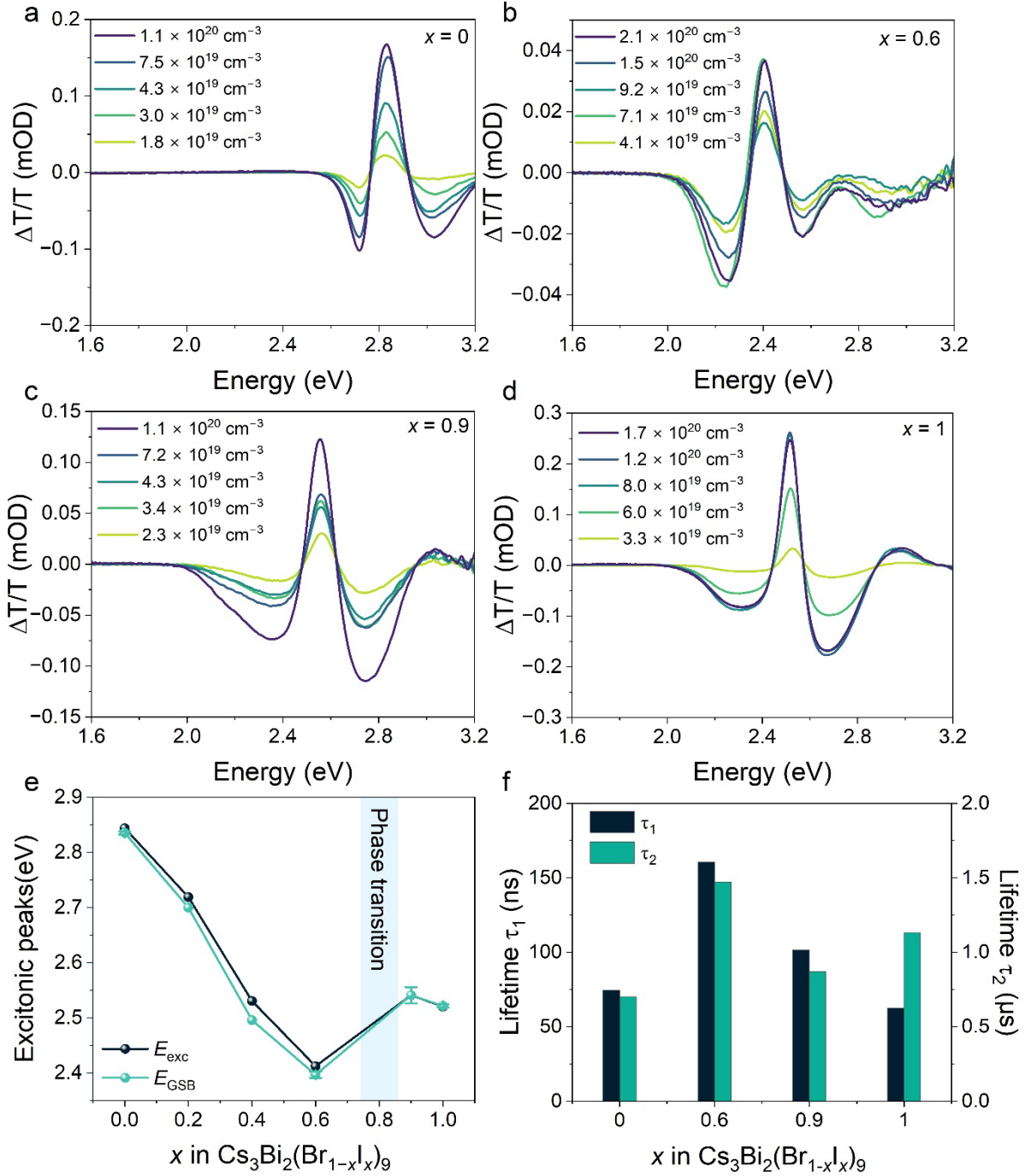

**Figure 4**: Charge carrier dynamics of $Cs_3Bi_2(Br_{1-x}I_x)_9$ thin films. Fluence-dependent transient transmittance $\Delta T/T$ spectra at delay time t = 23 ns of $Cs_3Bi_2(Br_{1-x}I_x)_9$ films with a) $x = 0$, b) $x = 0.6$, c) $x = 0.9$, and d) $x = 1$ after 355 nm laser pulse excitation. e) Peak energy of the steady-state excitonic peaks and transient ground-state bleach (GSB) peaks, f) The two decay time constants $\tau_1$ and $\tau_2$ of GSB for $Cs_3Bi_2(Br_{1-x}I_x)_9$ films with $x = 0$, $x = 0.6$, $x = 0.9$, and $x = 1$, observed in the timeframe of our set-up (1 ns to 400 µs).

To investigate the excited state dynamics of $Cs_3Bi_2(Br_{1-x}I_x)_9$ films, we used nanosecond transient absorption spectroscopy, which provides information on timescales relevant for long-lived excited states and trap states.[40] Prior to analysis, fluence-dependent experiments were performed to determine the optimum excitation power and investigate the photostability of the $Cs_3Bi_2(Br_{1-x}I_x)_9$ films. In our experiments, the initial carrier density is estimated from the absorption and thickness of the films,[41] and the calculated values are between $10^{19}$ to $10^{20}$ cm$^{-3}$. During the measurements with different pump powers, no change in the shape of the transient spectrum was noticed for the full bromide, iodide and mixed halides with $x = 0.6$ and 0.9 (**Figure 4**a-d), suggesting that these are stable under the different pump powers. By delaying the probe pulse with respect to the pump pulse, we obtained the time-dependent bleaching spectra shown in Figure S7. Note that there are no significant spectral changes except for the intensity during the decay. The spectra monitored at different time intervals after the laser pulse exhibits essentially the same features. These results indicate that the spectral shape is invariant as a function of time and the decay pathway remains the same.[42] We note that the bleach peak is red-shifted with increasing I content, aligning with the lower band gaps observed in the UV-vis absorption spectra. All ground-state bleaching (GSB) peaks (*i.e.*, positive value of ΔT/T) are located at almost the same excitonic peak positions (Figure 4e) as in the steady-state absorption spectrum, implying these features are related to exciton absorption.

To gain further insight into the charge carrier lifetime, we utilize the bi-exponential function $y = A_1\exp(-t/\tau_1) + A_2\exp(-t/\tau_1) + y_0$ to fit the decay of the GSB, where $\tau_1$ and $\tau_2$ are the time constants, $A_1$ and $A_2$ prefactors that indicate the relative contribution of each decay pathway and $y_0$ is a constant. The faster component ($\tau_1$) is on the order of 60–160 ns with the largest contribution to the signal and the slower component ($\tau_2$) on the order of 0.7–1.5 μs (Figure S8). These values are close to the time constant ($\tau_1$ = ~280 ns and $\tau_2$ > 1 μs) of the GSB in a $CH_3NH_3PbI_{3-x}Cl_x$ film.[43] However, unlike $CH_3NH_3PbI_{3-x}Cl_x$, the observed lifetime cannot be ascribed to radiative recombination, as this implies strong photoluminescence: however we were not able to detect any photoluminescence in our $Cs_3Bi_2(Br_{1-x}I_x)_9$ thin films. Previous TA studies on sub-nanosecond timescale found incomplete GSB recovery in $Cs_3Bi_2Br_9$ and $Cs_3Bi_2I_9$[28,44–47], hinting towards a long-lived (>1 ns) component in the excited state. On basis of our ns-μs TA measurements, we observe two recombination pathways with timescales of ~ 100 ns and 1.5 μs. Similar to the related material $Cs_2AgBiBr_6$,[48,49] the decay kinetics are independent of the charge carrier density (in the range $1.8 \times 10^{19}$ to $1.1 \times 10^{20}$ cm$^{-3}$) for $Cs_3Bi_2Br_9$ (Figure S9). In combination with the poor photoluminescence, it seems likely that recombination in these materials is mediated by defects. Typically, nanosecond time constants may originate from the recombination of charge carriers in shallow defect traps; for instance, 200 ns for defective $WSe_2$,[50] while for microsecond time scale are associated with recombination of immobilized charge carriers in surface states.[41,48] Therefore, the two time constants here most likely correspond to the kinetics involving surface defects such as vacancies, point defects, as well as grain boundaries.[42] Future efforts should focus on elucidating the chemical origin of these defects. Based on its lowest bandgap, reduced exciton binding energy, and enhanced stability, the composition with $x = 0.6$ is of particular interest for further optimization toward (indoor) photovoltaic and photocatalytic applications.

## 3. Conclusion

In this work, $Cs_3Bi_2(Br_{1-x}I_x)_9$ semiconductors were synthesised via mechanochemical synthesis. Substituting $Br^-$ with $I^-$ enable continuous band gap tuning from 2.59 to ~1.9 eV. XRD and Raman analysis on $Cs_3Bi_2(Br_{1-x}I_x)_9$ powders confirms the atomic-level halide mixing and compositional control across the series, except at $x = 0.8$, where a structural transition emerges. Optical characterization in thin films reveals composition-dependent exciton binding energies, reaching a minimum at $x = 0.6$. Interestingly, this minimum exciton binding energy coincides with the lower band-gap value, the smallest Urbach energy and the longest lifetimes, collectively indicating the mixing range (around $x = 0.6$) with the most favorable optoelectronic characteristics. These results highlight the potential of mechanochemically engineered $Cs_3Bi_2(Br_{1-x}I_x)_9$ semiconductors for stable, high-performance photoconversion applications, while more efforts may then focus on further optimization of these materials toward lower defect densities.

## 4. Experimental section

### 4.1. Chemicals

Cesium iodide (CsI, > 99.0%, TCI Europe N.V.), Cesium bromide (CsBr, >99.0%, TCI Europe N.V.), Bismuth iodide ($BiI_3$, 99%, Sigma-Aldrich), Bismuth bromide ($BiBr_3$, ≥ 98%, Sigma-Aldrich), Dimethyl sulfoxide (DMSO, anhydrous, ≥ 99.9%, Sigma-Aldrich), Chlorobenzene (anhydrous, 99.8%, Sigma-Aldrich) were used as received.

### 4.2 Synthesis of $Cs_3Bi_2(Br_{1-x}I_x)_9$ powder

$Cs_3Bi_2(Br_{1-x}I_x)_9$ powder was prepared by mechanochemical synthesis. Specifically, desired amounts of precursors (CsI, CsBr, $BiI_3$, $BiBr_3$) and 3 zirconia balls (diameter: 10 mm) were added into a 10 mL stainless steel jar with an inside zirconia wall in a nitrogen-filled glovebox. The ball-to-reactant mass ratio was around 10. The mixture was milled by shaking with a Retsch MM400 Mixer Mill for 3 cycles (20-minute of grinding at 30 Hz and 10-minute of pause for cooling).

### 4.3 Film fabrication

Quart substrates (dimension: 15 mm × 15 mm × 1mm) were scrubbed with Fairy soap liquid and washed by deionized water, followed by sonication in acetone and isopropanol for 20 minutes, respectively. After drying with an air blow gun, the substrates were treated with UV-ozone for 30 minutes, transferred into glovebox immediately and then pre-heated at 70 °C on a hot plate for 5 minutes before deposition.

0.25 mmol of $Cs_3Bi_2(Br_{1-x}I_x)_9$ powder were dissolved into 1 mL of DMSO and heated at 70 °C for 1 hour under magnetic stirring, followed by filtrating with a 0.1 μm Nylon-polypropylene syringe filter. The solution obtained was kept at 70 °C. Then, 100 μL of the above solution were spin coated on the clean quartz substates at 3000 rpm for 30 seconds (ramping rate: 1500 rpm/s). 200 μL of chlorobenzene were dynamically dropped in the center of the substrates at 15 seconds. The substates were transferred onto a hotplate at 70 °C for 15 minutes and then annealed at 250 °C for 5 minutes.

### 4.4 Characterization

XRD patterns were collected on a Bruker D2 Phaser with a Cu Kα X-ray source (λ = 1.5406 Å) operating at 30 kV and 10 mA) with a step size of 0.02° and a dwell time of 0.5 s.

Ultraviolet-visible (UV-vis) spectroscopy was recorded using a Perkin Elmer Lambda 950S spectrophotometer equipped with an integrating sphere. For powders, the diffuse reflectance spectra were acquired using polytetrafluoroethylene (PTFE) as a reference. For thin films, the measurements are also in an integrating sphere to alleviate the errors caused by light scattering. The "transflectance" (transmittance plus reflectance, $F_{TR}$) and the transmittance ($F_T$) were measured directly. To obtain the $F_{TR}$ spectra, the thin films were mounted in a variable angle clip-mount sample holder under an angle of 8° and positioned at the center of the integrating sphere. The $F_T$ spectra were collected when the thin film was placed in front of the sphere. Then, the absorptance ($F_A$) and reflectance ($F_R$) can be calculated according to the following equation[51]:

$F_A = 1 - F_{TR}$

$F_R = 1 - F_A - F_T$

The absorption coefficient (α) can be estimated from

$$\frac{F_T}{1 - F_R} = e^{-\alpha L} \tag{1}$$

Where L is the thickness of the thin film.

Raman spectra were acquired from a Horiba Xplora Plus Raman Spectrometer with a 50x objective lens and a 785 nm excitation laser (1200 gr/mm grating). The laser power was set to 1% for powders and 10% for thin films with an exposure time of 30 seconds. The spectra were calibrated with a Si wafer with a Raman band centered at 520 cm$^{-1}$.

Transient absorption measurements were carried out using an EOS broadband pump-probe sub-nanosecond transient absorption spectrometer (Ultrafast Systems LLC). The pump beams (355 nm and 532 nm) were generated using harmonic generation from a Nd:YAG laser (1064 nm, 50 ps FWHM pulse duration and 1 kHz repetition rate). A Leukos STM-2-UV supercontinuum laser (Limoges, France) with a spectral wavelength ranging from 350 to 1800 nm at a 2 kHz repetition rate and a pulse width (FWHM) of < 1 ns was coupled into EOS to generate a continuous white light as a probe pulse. The probe beam was split into two portions, one to the sample chamber and the other as a reference beam to correct the pulse-to-pulse fluctuation. The pump-probe delay is electronically controlled and measured by a Pendulum CNT-90 timer/counter/analyzer (Pendulum Instruments). The excitation laser power was measured using an OPHIR NOVA II Laser Power Meter. To determine the powder of pump laser, a OPHIR Nova II Power Meter was used. The intensity of the excitation beam is adjusted with a series of 25 mm absorptive neutral density filters.

The initial photogenerated carrier density ($n_0$) is calculated according to the following equation:[41]

$$n_0 = \frac{P \times ND\,(\%) \times F_A}{(hc/\lambda) \times A \times d} \qquad (2)$$

Where $P$ is pump power, determined by Nova II laser power & energy meter (Ophir Optronics Solutions Ltd.). $ND$ (%) is the percentage of transmitted light by 25 mm absorptive neutral density filters (Thorlabs, Inc.). $F_A$ is the absorptance (fraction of absorbed photons at the excitation wavelength). $h$ is the Planck's constant, $c$ is the speed of light and $\lambda$ is the wavelength of pump laser. $A$ is the area of pump spot (diameter: ca. $3 \times 10^{-4}$ cm$^{-2}$)[52] and $d$ is the thickness of sample (in cm).

Cross-sectional scanning electron microscopy (SEM) imaging was measured using an FEI Helios NanoLab G3 UC microscope and measured at 15 keV and 0.2 nA. The thin films were sputtered a 4 nm Au with a 108 Cressington manual sputter coater before measurement and mounted in a vertical SEM sample holder for thin samples.

**Acknowledgements**

This work was supported by the Advanced Research Center Chemical Building Blocks Consortium (ARC CBBC) and Open Competition ENW-XS (grant OCENW.XS24.2.029). H. Z. thanks Hanya Spoelstra assisting with the Raman spectroscopy characterization and Hui Wang for the SEM measurements.

**Data Availability Statement**

The data that support the findings of this study are available from the corresponding author upon reasonable request.

Received: ((will be filled in by the editorial staff))
Revised: ((will be filled in by the editorial staff))
Published online: ((will be filled in by the editorial staff))**References**

[1] S. Han, S. Zhu, G. Fan, X. Huang, G. Zhao, R. Shi, Y. Huang, M. Wu, J. Li, Y. Guo, Y. Gao, S. Zhuang, F. Cao, X. Yu, D. Zhang, *Adv. Funct. Mater.* **2025**, *35*, 2421770.

[2] M. Leng, Y. Yang, K. Zeng, Z. Chen, Z. Tan, S. Li, J. Li, B. Xu, D. Li, M. P. Hautzinger, Y. Fu, T. Zhai, L. Xu, G. Niu, S. Jin, J. Tang, *Adv. Funct. Mater.* **2018**, *28*, 170444611.

[3] Z.-J. Bai, S. Tian, T.-Q. Zeng, L. Chen, B.-H. Wang, B. Hu, X. Wang, W. Zhou, J.-B. Pan, S. Shen, J.-K. Guo, T.-L. Xie, Y.-J. Li, C.-T. Au, S.-F. Yin, *ACS Catal.* **2022**, *12*, 15157.

[4] J. Lee, A. Kumar, H. Tüysüz, *Angew. Chem. Int. Ed.* **2024**, *63*, e202404496.

[5] B. Park, B. Philippe, X. Zhang, H. Rensmo, G. Boschloo, E. M. J. Johansson, *Adv. Mater.* **2015**, *27*, 6806.

[6] S. M. Jain, D. Phuyal, M. L. Davies, M. Li, B. Philippe, C. De Castro, Z. Qiu, J. Kim, T. Watson, W. C. Tsoi, O. Karis, H. Rensmo, G. Boschloo, T. Edvinsson, J. R. Durrant, *Nano Energy* **2018**, *49*, 614.

[7] X. Chen, M. Jia, W. Xu, G. Pan, J. Zhu, Y. Tian, D. Wu, X. Li, Z. Shi, *Adv. Opt. Mater.* **2023**, *11*, 2202153.

[8] G. K. Grandhi, G. Koutsourakis, J. C. Blakesley, F. De Rossi, F. Brunetti, S. Öz, A. Sinicropi, M. L. Parisi, T. M. Brown, M. J. Carnie, R. L. Z. Hoye, P. Vivo, *Nat. Rev. Clean Technol.* **2025**, *1*, 132.

[9] S. Valastro, S. Gavranovic, I. Deretzis, M. Vala, E. Smecca, A. La Magna, A. Alberti, K. Castkova, G. Mannino, *Adv. Opt. Mater.* **2024**, *12*, 2302397.

[10] B.-B. Bin Yu, M. Liao, J. Yang, W. Chen, Y. Zhu, X. Zhang, T. Duan, W. Yao, S.-H. H. Wei, Z. He, *J. Mater. Chem. A* **2019**, *7*, 8818.

[11] S. Bonomi, P. Galinetto, M. Patrini, L. Romani, L. Malavasi, *Inorg. Chem.* **2021**, *60*, 14142–14150.15

## Supporting Information

**Tuning Excitonic Properties and Charge Carrier Dynamics by Halide Alloying in Cs$_3$Bi$_2$(Br$_{1-x}$I$_x$)$_9$ Semiconductors**

*He Zhao\*, and Eline M. Hutter\**

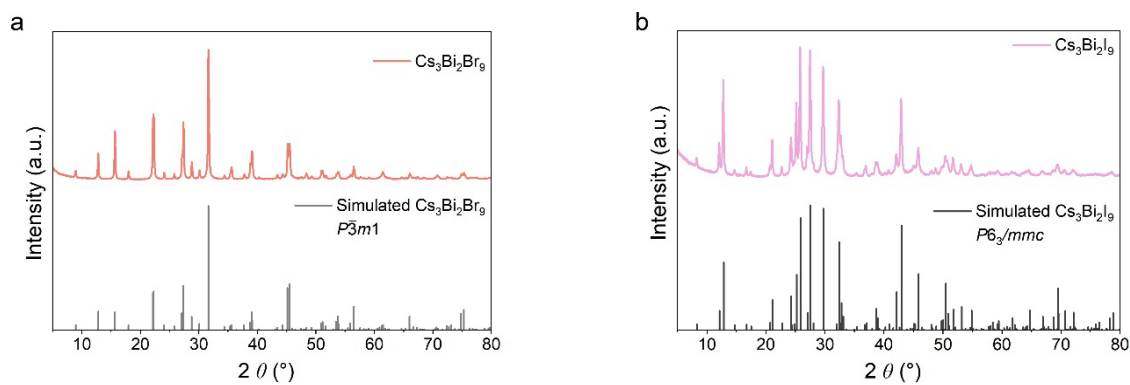

**Figure S1**. Powder XRD of a) $Cs_3Bi_2Br_9$ and b) $Cs_3Bi_2I_9$. The grey lines indicate the simulated XRD diffraction peaks for $Cs_3Bi_2Br_9$ and $Cs_3Bi_2I_9$ obtained by using the VESTA software. The crystallographic information files (CIF) are obtained from the crystallographic open database (COD). The COD ID for $Cs_3Bi_2Br_9$ is 2106376 and the COD ID for $Cs_3Bi_2I_9$ is 2106275.

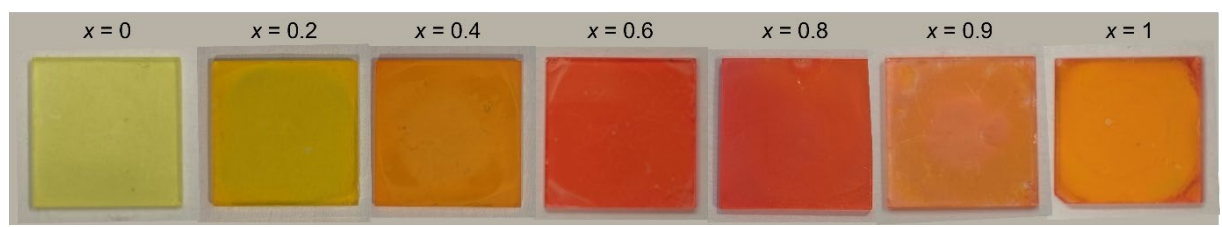

**Figure S2**. Optical images of $Cs_3Bi_2(Br_{1-x}I_x)_9$ thin films.

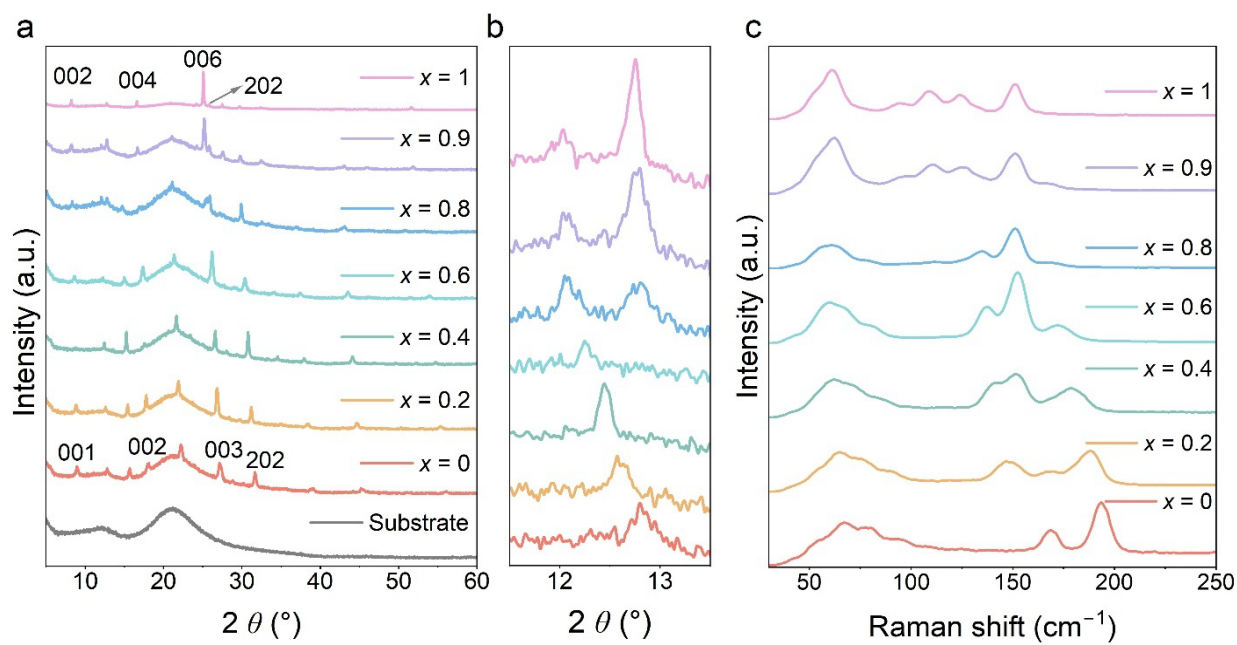

**Figure S3**. a) XRD patterns and b) enlarged view in a $2\theta$ range of $Cs_3Bi_2(Br_{1-x}I_x)_9$ thin films and quartz substrate. c) Raman spectra of $Cs_3Bi_2(Br_{1-x}I_x)_9$ thin films.

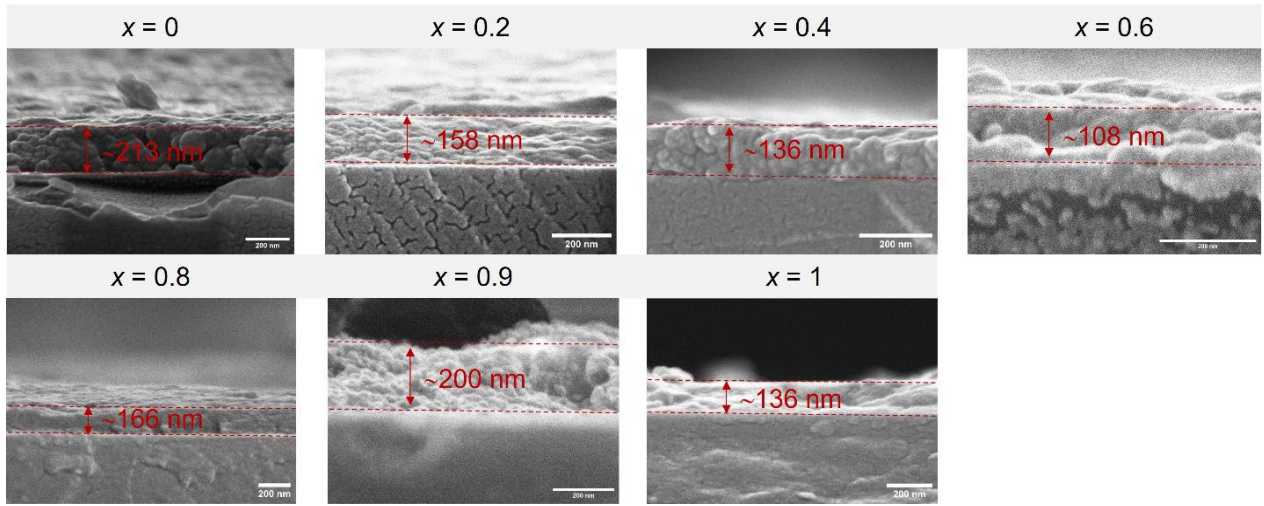

**Figure S4**. Cross-sectional SEM images of $Cs_3Bi_2(Br_{1-x}I_x)_9$ thin films.

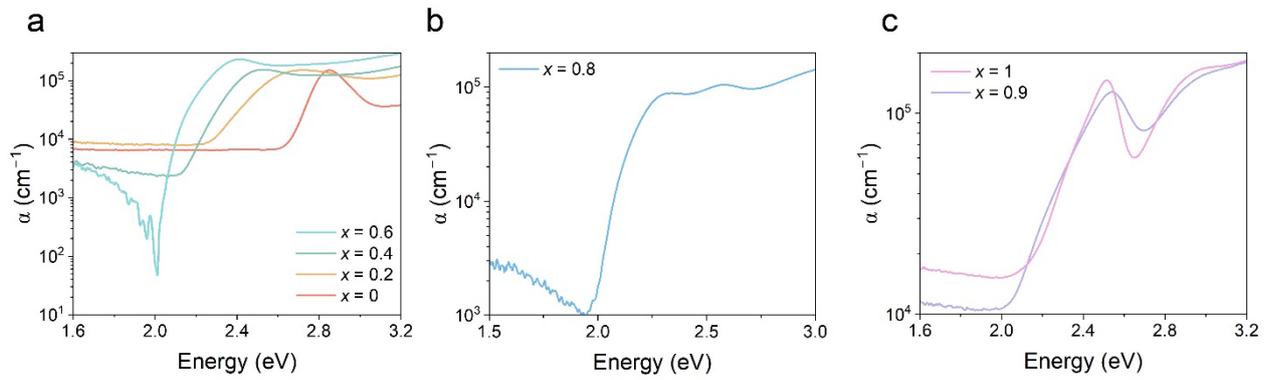

**Figure S5**. Plots of the measured absorption coefficient on a logarithmic scale, a) $x = 0-0.6$, b) $x = 0.8$, and c) $x = 0.9-1$.

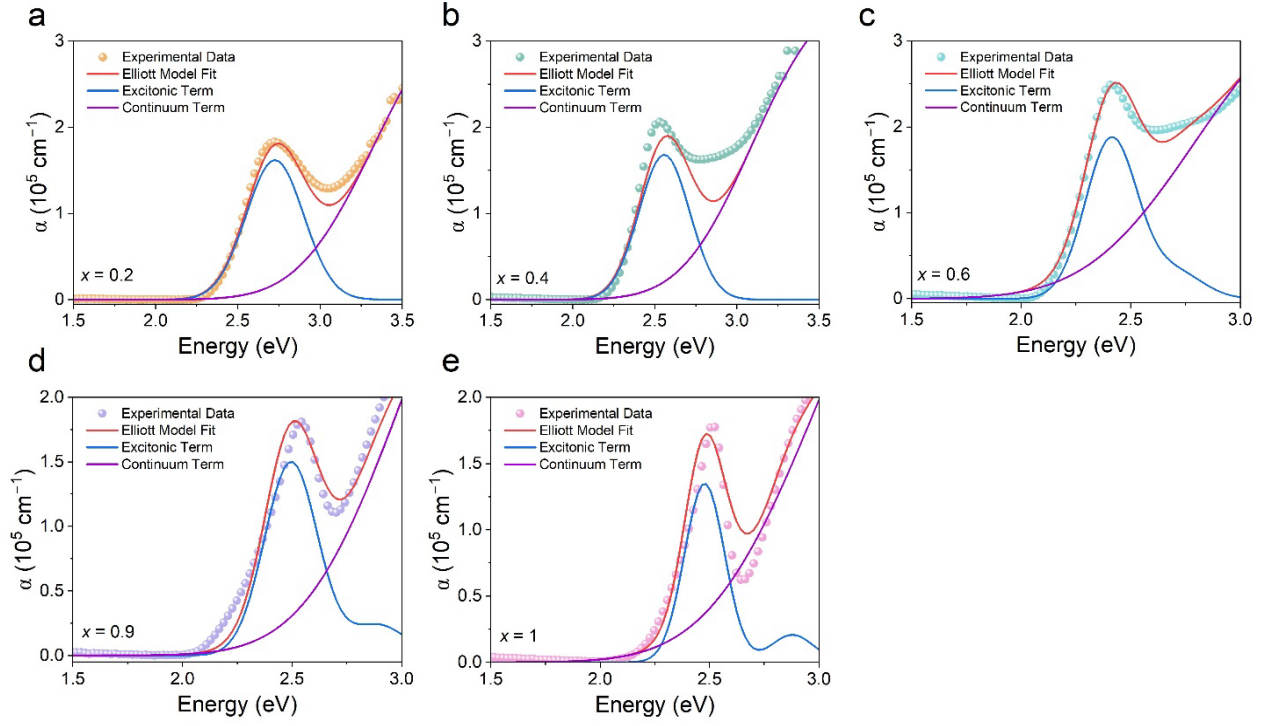

**Figure S6**. Absorption coefficient of a) $x = 0.2$, b) $x = 0.4$, c) $x = 0.6$, d) $x = 0.9$ and e) $x = 1$ of $Cs_3Bi_2(Br_{1-x}I_x)_9$ thin films fitted by Elliott's model (red line) with the contribution from excitonic (blue line) and continuum band (purple line) transitions.

**Table S1**. Fitted factors of $A_c$, $\lambda$, $\beta$, $\gamma$, $\delta$ and $E_G$ used in Elliot's model for extracting $E_B$ in $Cs_3Bi_2(Br_{1-x}I_x)_9$ thin films. $E_B$ (the exciton binding energy) is calculated as $E_G$ (band gap) $- E_{ex}$ (excitonic peak).

| $x$ | $A_c$ | $\beta$ (eV) | $\gamma$ | $\delta$ (eV) | $E_G$ (eV) | $E_{ex}$ (eV) | $E_B$ (meV) |
|---|---|---|---|---|---|---|---|
| 0 | 1.80 | 2.72 | 1.80 | 0.12 | 3.62 | 2.84 | 780 |
| 0.2 | 1.80 | 1.62 | 1.92 | 0.25 | 3.33 | 2.72 | 610 |
| 0.4 | 1.80 | 1.68 | 2.08 | 0.22 | 3.09 | 2.53 | 560 |
| 0.6 | 1.80 | 1.97 | 2.02 | 0.19 | 2.79 | 2.41 | 380 |
| 0.9 | 2.00 | 1.50 | 1.80 | 0.17 | 2.97 | 2.54 | 430 |

| 1 | 1.80 | 1.50 | 1.71 | 0.12 | 2.99 | 2.52 | 470 |

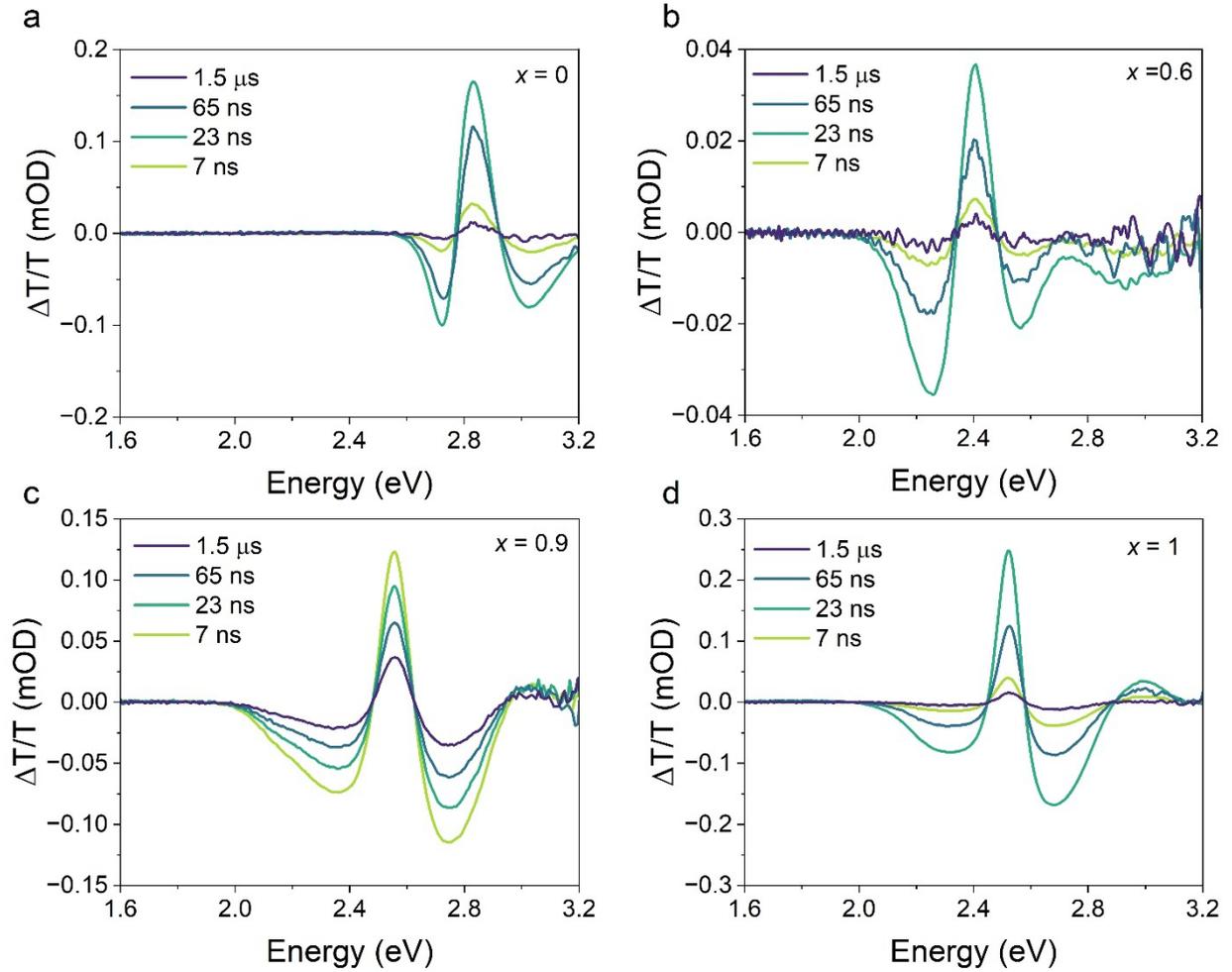

**Figure S7.** Transient transmittance ΔT/T spectra of a) $x = 0$ ($n_0$: $1.1 \times 10^{20}$ cm$^{-3}$), b) $x = 0.6$ ($n_0$: $2.1 \times 10^{20}$ cm$^{-3}$), c) $x = 0.9$ ($n_0$: $1.1 \times 10^{20}$ cm$^{-3}$) and d) $x = 1$ ($n_0$: $1.7 \times 10^{20}$ cm$^{-3}$) at different probe delay times following 355 nm laser excitation.

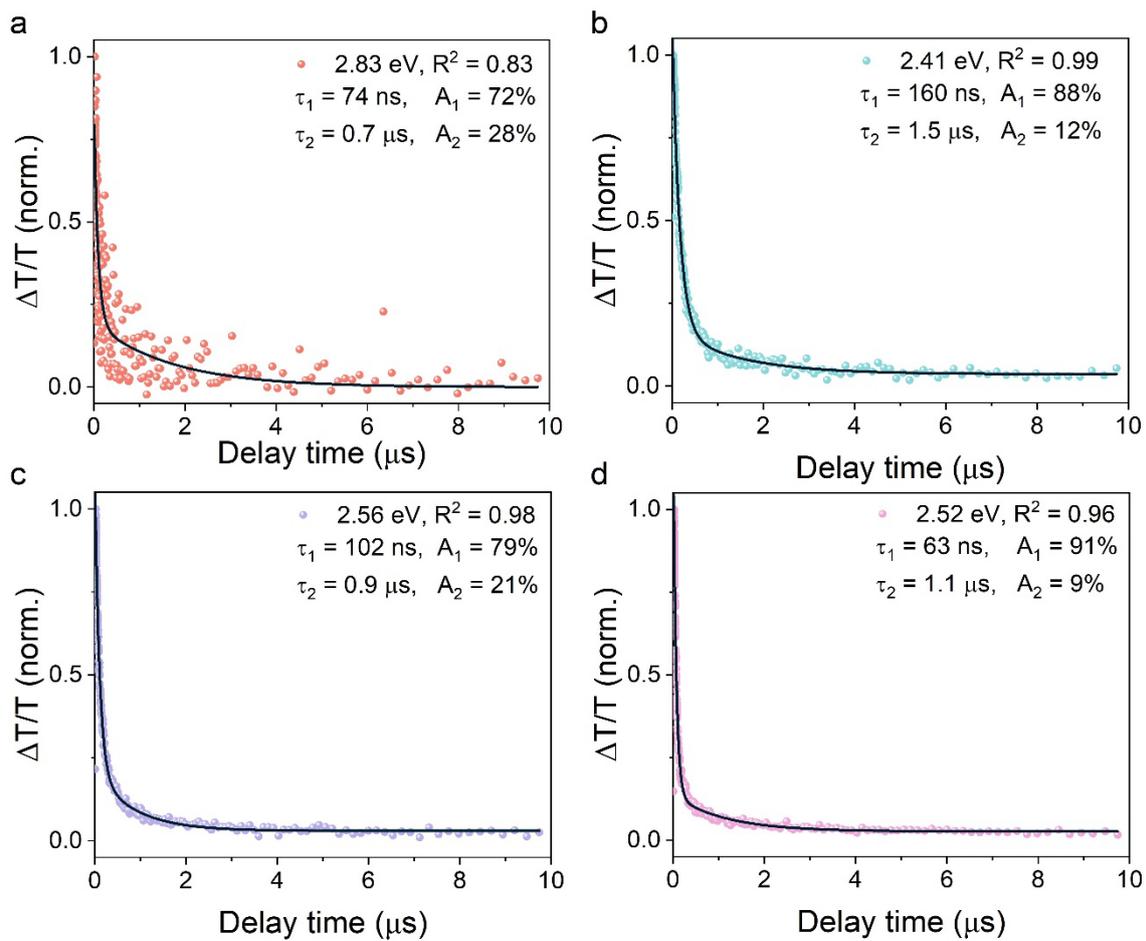

**Figure S8.** Bi-exponential fitting with $y = A_1\exp(-t/\tau_1) + A_2\exp(-t/\tau_1) + y_0$ at GSB of a) $x = 0$, b) $x = 0.6$, c) $x = 0.9$ and d) $x = 1$.

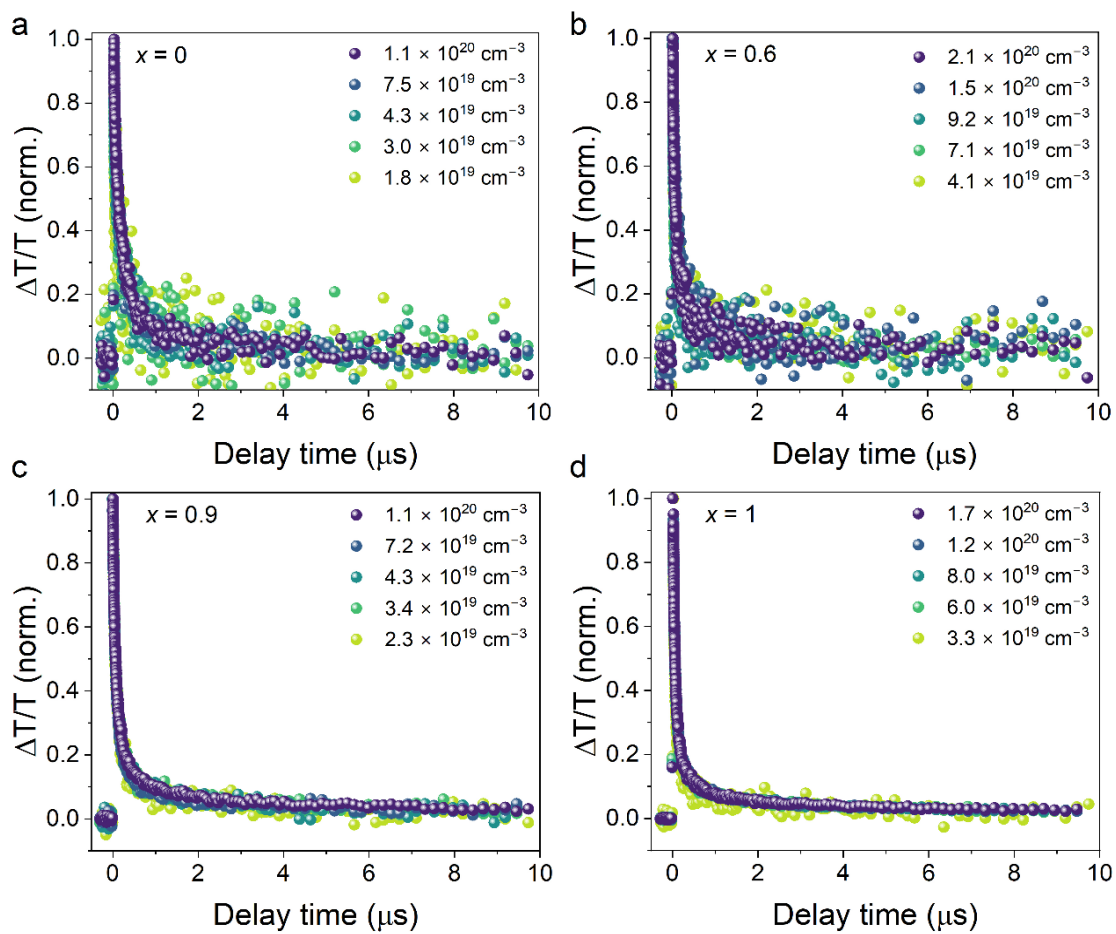

**Figure S9.** Fluence-dependent GSB decays of a) $x = 0$ (2.83 eV), b) $x = 0.6$ (2.41 eV), c) $x = 0.9$ (2.56 eV) and d) $x = 1$ (2.52 eV).